# Journal of Reinforced Plastics and Composites

**Compatibilization of HDPE/agar biocomposites with eutectic-based ionic liquid containing surfactant**

AA Shamsuri, R Daik, ES Zainudin and PM Tahir

The online version of this article can be found at:
http://jrp.sagepub.com/content/33/5/440

Published by:

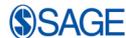

http://www.sagepublications.com

Additional services and information for *Journal of Reinforced Plastics and Composites* can be found at:

**Email Alerts:** http://jrp.sagepub.com/cgi/alerts

**Subscriptions:** http://jrp.sagepub.com/subscriptions

**Reprints:** http://www.sagepub.com/journalsReprints.nav

**Permissions:** http://www.sagepub.com/journalsPermissions.nav

**Citations:** http://jrp.sagepub.com/content/33/5/440.refs.html

>> Version of Record - Mar 11, 2014

OnlineFirst Version of Record - Jan 27, 2014

What is This?

# Compatibilization of HDPE/agar biocomposites with eutectic-based ionic liquid containing surfactant

**AA Shamsuri[1,2], R Daik[2], ES Zainudin[1] and PM Tahir[1]**

## Abstract
In this research, eutectic-based ionic liquid specifically choline chloride/glycerol was prepared at a 1:2 mole ratio. The choline chloride/glycerol was added with the different content of surfactant (hexadecyltrimethylammonium bromide). The choline chloride/glycerol-hexadecyltrimethylammonium bromide was introduced into high-density polyethylene/agar biocomposites through melt mixing. The mechanical testing results indicated that the impact strength and tensile extension of the biocomposites increased with the introduction of the choline chloride/glycerol-hexadecyltrimethylam-monium bromide. The scanning electron microscope, differential scanning calorimetry and thermal gravimetric analysis results exhibited that significant decrease in the number of agar fillers pull-out, melting point and thermal decomposition temperatures of the biocomposites are also due to the choline chloride/glycerol-hexadecyltrimethylammonium bromide. The Fourier transform infrared spectra and X-ray diffractometer patterns of the biocomposites introduced with the choline chloride/glycerol-hexadecyltrimethylammonium bromide demonstrate the presence of physical interactions, which contributes to the increase of compatibility between both high-density polyethylene and agar. In conclusion, high-density polyethylene/agar biocomposites could be compatibilized with eutectic-based ionic liquid containing surfactant, choline chloride/glycerol-hexadecyltrimethylammonium bromide.

## Keywords
High-density polyethylene, agar, biocomposite, choline chloride/glycerol, hexadecyltrimethylammonium bromide

## Introduction

The advancement of the polymer composites industry has been very encouraging in the past years. The polymer biocomposites have also received considerable attention, this is most probably because the production cost of the composite products can be lowered through consumption of natural fibers as fillers.[1,2] Moreover, the utilization of synthetic polymers and synthetic fibers can also be reduced and this indirectly can conserve the environment.[3,4] Furthermore, the biocomposites that are fabricated by using polyethylene as a matrix could have the positive implications since this polymer typically possesses a quantity of feasible characteristics such as biocompatible, non-hazardous and chemical-resistant.[5] In addition, polyethylene for example high-density polyethylene (HDPE) is inexpensive, has high tensile stress (compared to low-density polyethylene), is easy for processing and is recyclable as well.[6] HDPE also has low melting point (in comparison with polypropylene), it is thus requiring low energy consumption for processing of biocomposites and avoiding thermal decomposition of the natural fibers.

Agar is a polysaccharide that consists of a mixture of two components, agarose and agaropectin (see Figure 1(a)), in variable fractions depending on the original raw material. The use of agar as fillers for

[1]Laboratory of Biocomposite Technology, Institute of Tropical Forestry and Forest Products, Universiti Putra Malaysia, Selangor, Malaysia
[2]School of Chemical Sciences and Food Technology, Faculty of Science and Technology, Universiti Kebangsaan Malaysia, Selangor, Malaysia

**Corresponding author:**
ES Zainudin, Laboratory of Biocomposite Technology, Institute of Tropical Forestry and Forest Products, Universiti Putra Malaysia, 43400 UPM Serdang, Selangor, Malaysia.
Email: edisyam@upm.edu.my

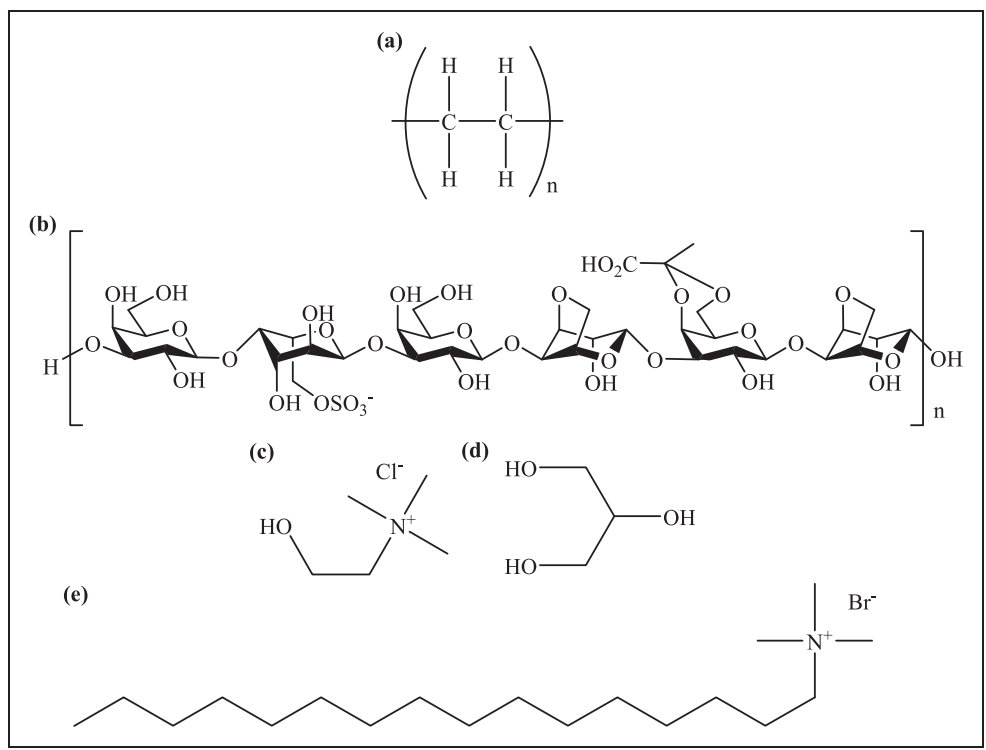

**Figure 1.** Chemical structures of the (a) HDPE, (b) agar, (c) ChCl, (d) Gly, and (e) HTAB.

various polymer composites systems have been widely investigated by many researchers, because it is abundant in nature, low-cost, non-toxic, environment friendly, renewable and has gelling ability as well.[7,8] Besides, the produced HDPE/agar biocomposites may also be used in production of biocompatible and biodegradable characteristics products. Anyhow, the compatibility between the HDPE and agar in the biocomposites system is very low, thus resulting in significant reductions of their performance, particularly the mechanical properties (impact strength, tensile extension, tensile strain, etc.) of the final resulting composite product. The incompatibility is because the HDPE is a thermoplastic polyolefin which is normally non-polar, hydrophobic and has low free energy surface.[9,10] Conversely, the agar is a biopolymer which is naturally polar, hydrophilic and has high free energy surface[11,12] and therefore the importance of compatibilization of the biocomposites could not be neglected.

In response to this, the chemical treatments by using alkalis, acids, coupling agents or compatibilizers were usually carried out to compatibilize between two different components especially for a better adhesion at their interface and also for improving physicomechanical properties of biocomposites, as demonstrated by many researchers.[13–16] Nonetheless, some of these treatments have negative effects since they are predominantly consumed hazardous chemicals as well as required more energy, time and additional steps.

In recent times, ionic liquids were effectively used as compatibilizers in polymeric composites system specifically for hydrophobic polymer and hydrophilic filler.[17] Moreover, in our previous study we have also demonstrated that the ionic liquid when mixed together with urea could be used as a coupling agent for agarose/talc composite films, the results showed that the ionic liquid/urea mixture improved interfacial adhesion between agarose and talc particles.[18] Additionally, the use of ionic liquids also seems fascinating because they are environmentally benign solvents and display some exceptional features, for instance can dissolve or solvate many organic and inorganic (liquids/solids) chemical compounds including a number of polymers and biopolymers.[19–22] Ionic liquids are chemically inert, thermally stable, biocompatible, and most importantly, they can be modified according to the needs, nonflammable, non-volatile and also very promising replacements for the traditional toxic volatile organic solvent.[23–26]

In the present study, we have used eutectic-based ionic liquid since it frequently owns some practical properties, for instance easy to prepare, physical as well as chemical properties are similar to the common ionic liquids. Moreover, its precursor can also be easily found, is of low cost, non-hazardous and biodegradable. Among several eutectic-based ionic liquids, choline chloride/glycerol (ChCl/Gly) has some valuable characters, for example very low freezing point ($-40^\circ$C), eco-friendly, cheap and biocompatible.[27] ChCl/Gly utilized in this study was added with surfactant to enhance compatibility between HDPE and agar. This is because the surfactants are amphiphilic compounds that contain polar and non-polar parts (hydrophilic and hydrophobic alkyl groups),[28,29] therefore they induce the polymer composites components to interact with each other.[30,31] Cationic surfactant, specifically hexadecyltrimethylammonium bromide (HTAB), was selected for exploration since its molecules were able to provide intermolecular interactions between polar and non-polar polymers as reported in our preceding study.[32] These could increase compatibility between the components of the biocomposites. In addition, the use of ionic liquid is also to solvate surfactant[33] so as to facilitate HTAB to interact with both HDPE and agar in the biocomposites.

So far, the study concerning the polyolefin/polysaccharide biocomposites that were prepared with the introduction of eutectic-based ionic liquid containing surfactant has not been reported. Thus, the purpose of the research is to investigate the possibility of ChCl/Gly-HTAB to compatibilize the biocomposites especially comprising HDPE and agar. The mechanical, morphological, thermal and interactional properties of the HDPE/agar biocomposites have been characterized, studied and comprehensively discussed further in detail.

## Materials and method

### Materials

The eutectic-based ionic liquid precursor used is a choline chloride, denoted as ChCl (purity 99%) and obtained from Acros. Glycerol, Gly (purity 99%) as a hydrogen bond donor was also purchased from Acros. The surfactant used is a hexadecyltrimethylammonium bromide, abbreviated as HTAB (purity 99%) and procured from Sigma Aldrich. The polymeric matrix used is a high-density polyethylene, HDPE (medium molecular weight distribution grade), acquired from Etilinas Polyethylene Malaysia Sdn. Bhd. The filler of biocomposites is an agar powder, 150 μm (plant culture grade), obtained from Sigma Aldrich. All the chemicals were used as received without further purification and treatment. The chemical structures of materials that have been used in this research are shown in the Figure 1.

### Preparation of eutectic-based ionic liquid containing surfactant

The ChCl/Gly eutectic-based ionic liquid was made via complexation reaction[34,35] in accordance with our earlier publication or can also be prepared as described by other authors.[36,37] The ChCl was mixed with Gly at a mole ratio of 1:2 in a beaker, heated at temperature of 100°C and stirred rigorously by using a magnetic stirrer hot-plate apparatus until a transparent homogeneous liquid was obtained. The eutectic-based ionic liquid containing surfactants were then prepared separately in a closed glass bottle by adding the HTAB into ChCl/Gly. After that, they were stirred at 100°C until clear liquid was acquired. The prepared ChCl/Gly containing HTAB (ChCl/Gly-HTAB) was then stored in an oven at 60°C before use.

### Preparation of HDPE/agar biocomposites

The biocomposites were prepared through melt mixing by using a Brabender internal mixer equipped with real-time processing torque recorder. The mixing was carried out at a temperature of 150°C, the rotor speed was fixed at 40 r/min. The HDPE (24 g) was charged into the Brabender mixing chamber at the beginning for 3 min followed by agar (16 g) for 3 min and ChCl/Gly (4.44 g) or ChCl/Gly-HTAB for 3 min. Then, the mixing time was continued for another 3 min where plateau torque was achieved. The duration of the whole process was 12 min. The mixed biocomposites obtained from the internal mixer were converted into 1-mm sheet using a compression-molding machine. The molding procedures involved preheating at 150°C for 7 min and followed by 2 min of compression at the same temperature and subsequent cooling under pressure for 5 min. The resultant biocomposites were conditioned in an oven at 80°C for at least up to 24 h prior to characterizations. The HDPE and agar weight ratio was constant and fixed at 60/40, whereas the composition between HDPE/agar and ChCl/Gly loading was stipulated at 90/10 (wt/wt), while the weight percentages of HTAB were varied from 0 to 1.0 wt% relative to biocomposites content. For the sake of comparison, the biocomposite containing only HDPE and agar has also been prepared. The weight proportion of the HDPE/agar biocomposites with and without introduction of ChCl/Gly and ChCl/Gly-HTAB is summarized in Table 1.

### Characterization

*Mechanical testing.* The impact strength of the biocomposites was measured according to the ASTM D4812[38] by using a CEAST impact testing machine (model 9050) via Izod impact test method. The unnotched sample was supported by a vertical cantilever beam and was broken by a single swing of the 0.5 Joule pendulum. The samples for the impact tests were cut using a table saw machine into the form of rectangular bars with nominal size of $60 \times 12.7 \times 1.0\,\text{mm}^3$.[39]

**Table 1.** The weight proportion of the HDPE/agar biocomposites with and without introduction of ChCl/Gly and ChCl/Gly-HTAB.

| Sample | Weight proportion | | |
|---|---|---|---|
| | HDPE/Agar (60/40) | ChCl/Gly | HTAB |
| A | 100 | 0 | 0 |
| B | 90 | 10 | 0 |
| C | 90 | 10 | 0.2 |
| D | 90 | 10 | 0.6 |
| E | 90 | 10 | 1.0 |

HDPE: high-density polyethylene; ChCl/Gly: choline chloride/glycerol; HTAB: hexadecyltrimethylammonium bromide.

The measurements were conducted at room temperature (25°C). Ten samples were used for each composition and the average data along with corresponding standard deviation were calculated.

The tensile extension, tensile stress and modulus properties were ascertained according to the ASTM D638[40] at room temperature by using an Instron universal testing machine (model 5567) equipped with a 30 kN load cell. The crosshead speed was 5 mm min$^{-1}$ with a 50-mm gauge length. The biocomposites samples of 1 mm thickness were cut into a dumbbell shape using a CEAST cutter. The mean value, taken from ten samples for each composition, was reported with the standard deviation to show the error range.

*Scanning electron microscope.* The morphology of the biocomposites was viewed by using scanning electron microscope (SEM) (JEOL, model JSM-6400). The samples for morphological examination were cross-sectioned by making an impact fracture at room temperature and they were also gold-coated. The morphologies were observed at magnification of ×100 with accelerating voltages of 15 kV.

*Differential scanning calorimetry.* The melting point temperature, $T_m$, value of the biocomposites was ascertained with a differential scanning calorimetry (DSC) apparatus (Mettler Toledo DSC822$^e$), under a constant stream of compressed air at a flow rate of 50 ml min$^{-1}$. The samples were tightly sealed in aluminium crucibles and first heated at 100°C with an isotherm of 7 min to eliminate the thermal history. The analysis was done over a temperature range of 40 to 200°C at a heating rate of 10°C min$^{-1}$.

*Thermal gravimetric analysis.* The thermal decomposition behaviour of the biocomposites was studied by means of thermal gravimetric analysis (TGA; Mettler Toledo TGA/SDTA 851$^e$). The analyses were operated with heating rate of 10°C min$^{-1}$ and the temperature ranged from 40 to 600°C. The samples were heated in the alumina pans in an atmosphere of compressed air with flow rate of 50 ml min$^{-1}$.

*Fourier transform infrared.* The functional groups and the interactions between the components of the biocomposites were determined by using Fourier transform infrared (FTIR; Perkin Elmer Spectrum 100 Series). The FTIR spectra were obtained by employing a universal attenuated total reflectance (UATR) equipped with a ZnSe-diamond composite crystal accessory. Each sample was scanned 16 times, in the frequency range of 4000 to 500 cm$^{-1}$ and resolution of 4 cm$^{-1}$ at ambient temperature.

*X-ray diffractometer.* The crystallographic properties of the biocomposites were characterized by employing an X-ray diffractometer (XRD; PANalytical Philips, X'Pert PRO MPD PW3040) equipped with a monochromatized radiation source of CuKα of 1.5405 Å. The voltage and current intensities were 40.0 kV and 30.0 mA, respectively. All samples were scanned in the range of 20 to 25° 2θ with a step size of 0.01° and step time of 0.2 s at room temperature.

## Results and discussion

### Melt processing characteristics

The processing torque-time curves of the HDPE/agar biocomposites with different contents of ChCl/Gly-HTAB are demonstrated in the Figure 2. The sharp increase in the processing torque is acquired for all samples at around the first minute as a result of the increasing resistance exerted on the internal mixer rotors by the unmelted HDPE during mixing process. The peaks start to decrease with increasing of mixing time as the melting of HDPE takes place. The processing torque values began to rise again to high values immediately at around the fourth minute as can be seen in all samples after the charge of agar. This is because more forces are needed to disperse the agar fillers in the molten HDPE. The torque starts to decrease as the agar becomes very well dispersed into the matrix. For sample A, as the filler dispersion is completed, the torque value decreases and remains almost constant at a particular level until the end of the total mixing time. At around the seventh minute, at which ChCl/Gly or ChCl/Gly-HTAB was charged into the mixer, a downward peak is recorded. The slight, sudden drop in torque value for sample B represents the sliding effect of the ChCl/Gly. Interestingly, the samples C to E show very broad downward peaks as a result of the lubricating effect of the ChCl/Gly-HTAB. As dissemination of ChCl/Gly or ChCl/Gly-HTAB is completed, the torque started to increase gradually, due to the increase in biocomposites melt viscosity, and to settle at a more stable value.

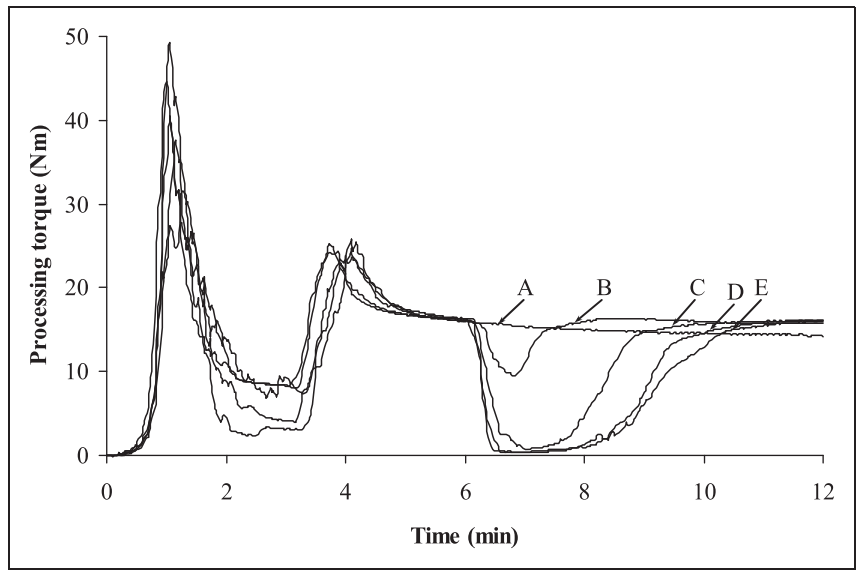

**Figure 2.** Processing torque–time curves of HDPE/agar biocomposites with different contents of ChCl/Gly-HTAB.

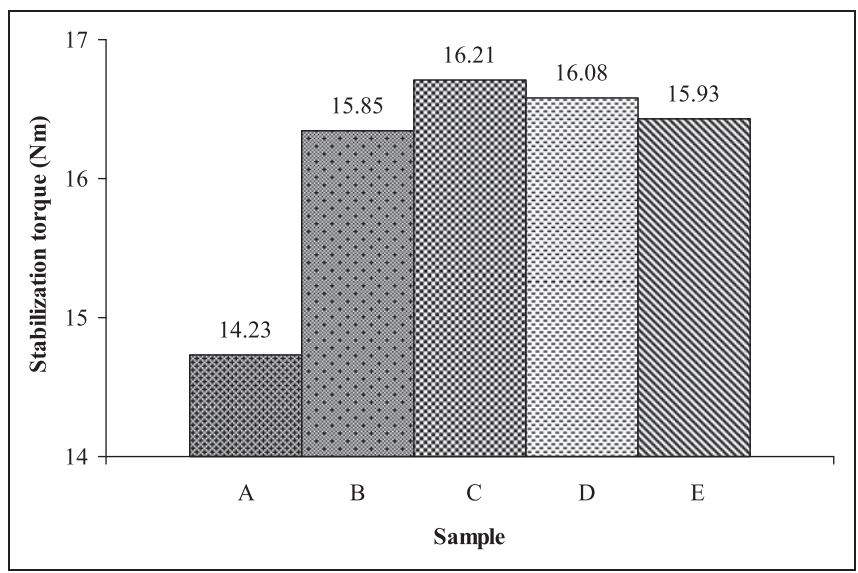

**Figure 3.** Effects of ChCl/Gly and ChCl/Gly-HTAB loadings on the stabilization torque of the HDPE/agar biocomposites.

The effects of ChCl/Gly and ChCl/Gly-HTAB loadings on the stabilization torque of the HDPE/agar biocomposites are shown in the Figure 3. The stabilization torque values are attained from the end time of the mixing process (at twelfth minute). The different stabilization torque values were obtained for all the cases. This is largely related to the distinct loading of ChCl/Gly and ChCl/Gly-HTAB that was charged into the mixing chamber. In case of sample B, it is clearly seen that the stabilization torque increased with the introduction of ChCl/Gly. From our perspective, this increase is caused by the weight ratio between HDPE and agar fixed to 60/40, and their weight proportion also was fixed when ChCl/Gly was introduced. As a result, reducing the mobility of HDPE molecular chains and consequently increasing the stabilization torque value. On the other hand, the stabilization torque of sample C further increased with the introduction of ChCl/Gly-HTAB. However, it can be seen that the stabilization torque of samples D and E continuously decreases as the ChCl/Gly-HTAB loading in the biocomposites increased, but they are still higher than samples A and B. Therefore, the highest stabilization torque value was observed at sample C, followed by samples D and E. As demonstrated in this result, the biocomposites with ChCl/Gly-HTAB reflect possible better interactions than biocomposites with ChCl/Gly and without ChCl/Gly-HTAB at a similar HDPE/agar loading. It is owing to the addition of HTAB into the ChCl/Gly, which improves the matrix-filler interfacial adhesions between HDPE and agar and results in increase of viscosity of the molten biocomposites.[32,41] This certainly could be correlated with the increase of the compatibility between the components of biocomposites.

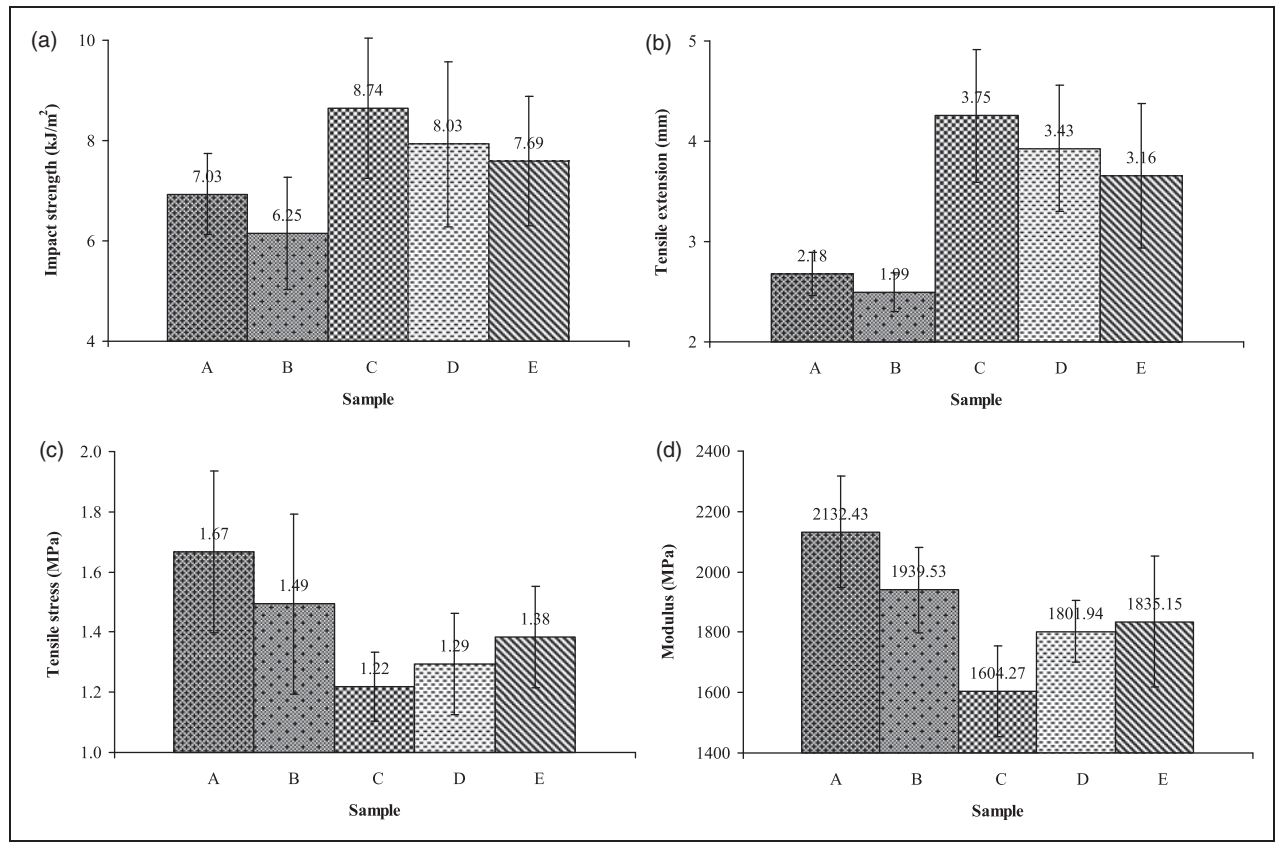

**Figure 4.** Effects of ChCl/Gly and ChCl/Gly-HTAB contents on the (a) impact strength, (b) tensile extension, (c) tensile stress, and (d) modulus at break of the HDPE/agar biocomposites.

### Mechanical properties

Impact strength and tensile extension at break of the HDPE/agar biocomposites are presented in Figure 4(a) and (b), respectively. It is clearly seen that the impact strength and tensile extension values of sample A were 7.03 kJ/m$^2$ and 2.18 mm, respectively, exhibiting low ductile behavior. It seemed that the applied impact on biocomposite cannot be easily transferred from HDPE matrix to agar fillers because of incompatibility between them. Sample B showed the lowest impact strength and tensile extension properties, where its values were only 6.25 kJ/m$^2$ and 1.99 mm, respectively. Although the impact and extension properties of the biocomposite are expected to be high with introduction of ionic liquid, ironically it is seen that both properties decreased as the ChCl/Gly was introduced. With deterioration of these properties, the results exhibit that the ChCl/Gly is not able to compatibilize between both HDPE and agar or even plasticize the biocomposites. Hence, it could not act either as compatibilizing or plasticizing agent in the biocomposites system. However, the addition of HTAB into the ChCl/Gly has improved the impact strength and tensile extension of biocomposites. From our point of view, the ChCl/Gly ionic liquid acts as a solvating agent by disintegrating the crystal lattices of the HTAB and complexing with their individual ions. The best compatibilization effect of the ChCl/Gly-HTAB on the HDPE/agar biocomposite was discovered in sample C with loading of 0.2 wt% HTAB. This is because at this level impact strength and tensile extension were found to be higher in comparison with other biocomposites samples. The impact strength of sample C showed about 24% increment and it also showed an improvement of the tensile extension at break up to 72%. The impact strength and tensile extension of the biocomposites samples D and E decrease continuously by increasing the ChCl/Gly-HTAB content. In many cases, the decreases in mechanical properties of polymer composites are due to the high content or extra amount of the surfactant.[42] At such high content of HTAB, they tend to form aggregates that act as a mild tough failure concentrator in the biocomposites. But, at lower content of HTAB, they act as a compatibilizing agent. Nevertheless, the decreases in impact and extension properties for samples D and E are still not beyond the properties of biocomposites samples A and B.

The tensile stress and modulus at break of the biocomposites are shown in the Figure 4(c) and (d), respectively. The reinforcing ability of agar in biocomposites with ChCl/Gly seems not very impressive. Thus, the introduction of ChCl/Gly in sample B not only reduced the impact strength and tensile extension of the biocomposites but also decreased their tensile stress and modulus. This is majorly corresponding to high mobility and high polarity of ChCl/Gly that can interact with agar fillers solely causing them become softer and weaker. On the other hand, biocomposite sample C exhibits the decreasing trend in tensile stress and modulus. This is absolutely the typical behavior of a ductile material of sample C. While, with further increase of ChCl/Gly-HTAB, the tensile stress and modulus of biocomposites samples D and E are slightly increased. As can be seen in these results, when comparing samples A and B with samples C to E, it was obvious that the tensile stress and modulus properties of the former were significantly greater than that of the latter.

## Morphological observation

The SEM micrographs of the fractured surfaces (internal sections) of the HDPE/agar biocomposites are displayed in the Figure 5. For the biocomposite sample A, as perceived in the Figure 5(a), a number of such holes (delineated with white dashed circles) are quite higher than the others due to the agar fillers pull-out. The SEM micrograph further shows the existence of voids around the fillers and there is no intimate contact between the HDPE and the agar. From this observation, it is clearly revealed that the agar fillers surfaces cannot be wetted by the molten HDPE matrix caused by the distinct compatibility as well as polarity, hydrophobicity and free energy surface properties, as we stated above. After the introduction of ChCl/Gly, the surface morphology of the biocomposite sample B is not considerably changed in terms of quantity of such holes, as observed in the Figure 5(b). Fillers pull-out are also traced in the biocomposite sample B, this is implying that the compatibility between the components of biocomposites was also very poor. Nevertheless, biocomposites samples C to E showed the differences in the SEM micrographs compared with the biocomposites samples A and B.

It can be seen in the Figure 5(c), when biocomposite introduced with ChCl/Gly-HTAB, the morphology of the fractured surface of the sample C has changed, whereby the number of such holes was diminished. Therefore, by the addition of just 0.2 wt% HTAB in the biocomposite greatly decreased the amount of agar fillers pull-out. On the other hand, biocomposites bearing higher ChCl/Gly-HTAB contents (>0.2 wt% HTAB) showed an insignificant increase in the agar fillers pull-out. Nonetheless, the adhesion of the agar fillers to the HDPE matrix is quite good though few such holes are seen in the biocomposites samples D and E, as shown in the Figure 5(d) and (e), respectively. This result clearly indicates that the introduction of the ChCl/Gly-HTAB enhances the interfacial adhesion between HDPE and agar in the biocomposites system. It is also found that there are little or no voids around the agar fillers of the biocomposites. The decline of the number of filler pull-out and voids attributed to the enhancement of compatibility between the biocomposites components.[43] Hence, these observations show that the compatibility between these components is increased. This is resulting from the presence of interactions between agar fillers and HDPE matrix. The findings from the SEM micrographs are in very good agreement with the impact strength as discussed in mechanical properties results in the section above.

## DSC characterization

The DSC thermograms of HDPE/agar biocomposites as a function of ChCl/Gly-HTAB loadings are indicated in Figure 6. The melting point temperature ($Tm$) values of the biocomposites are listed in Table 2. It is known that the glass transition temperature ($T_g$) value is always used to determine the miscibility and compatibility between blends components.[32] However, in this study we have found that the $T_g$ of the biocomposites could not noticeably be observed in the range of temperature studied. Their $T_g$ values are extremely low (usually below $-100°C$) due to the use of HDPE as a matrix.[44] Nonetheless, the $Tm$ value of the biocomposites can also give indirect information about the miscibility and compatibility between components in the polymer blends and composites systems.[11,45] The peaks obtained from DSC thermograms can be attributed to the specific $T_m$ of crystalline phase of the biocomposites.

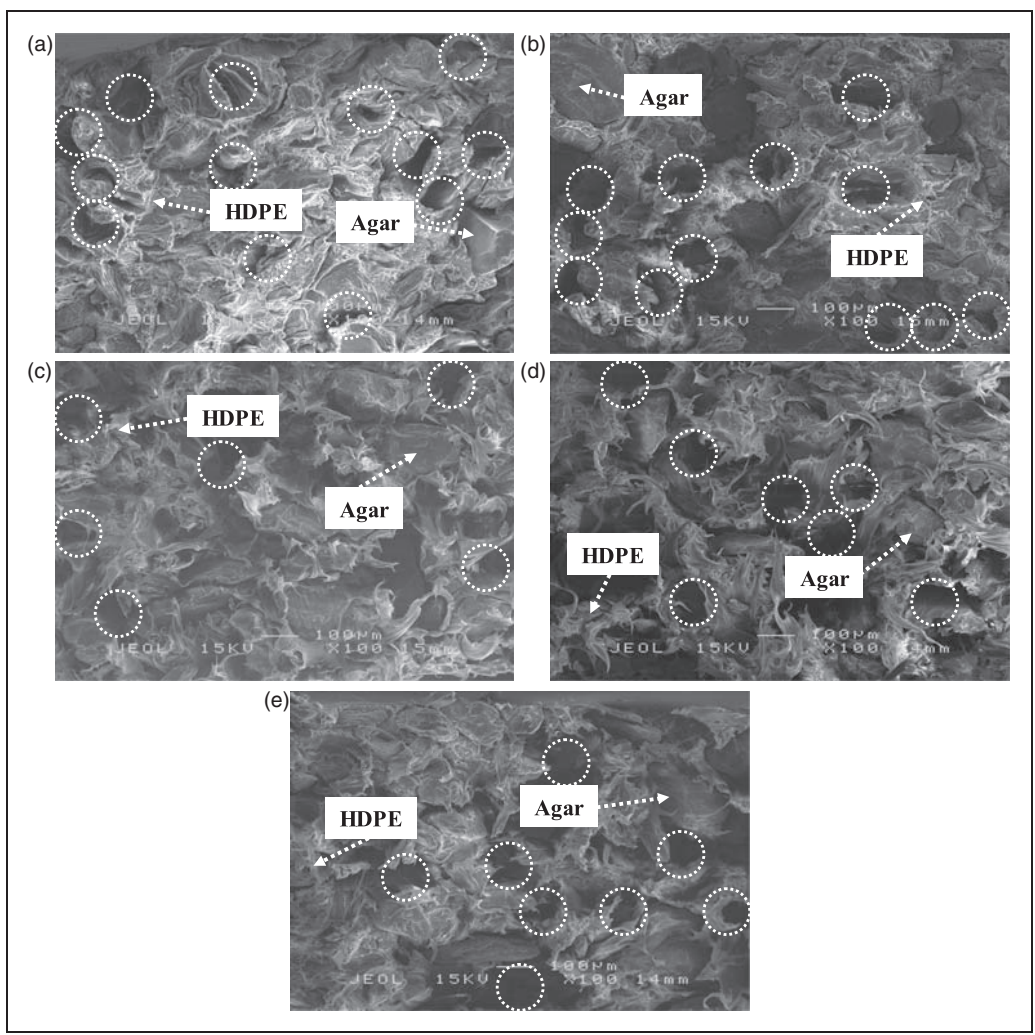

**Figure 5.** SEM micrographs of the fractured surfaces of the HDPE/agar biocomposites at magnification of 100×. HDPE: high-density polyethylene.

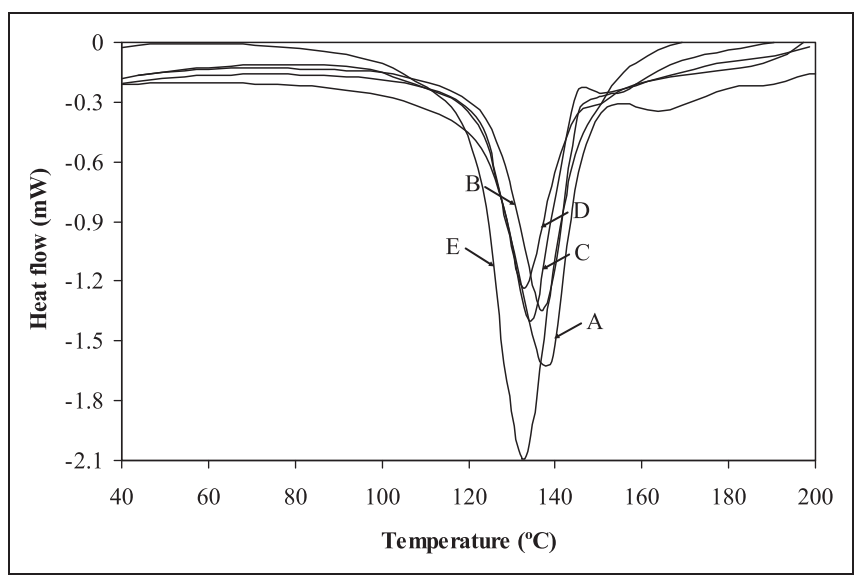

**Figure 6.** DSC thermograms of the HDPE/agar biocomposites as a function of ChCl/Gly-HTAB loadings.

Referring to Table 2, the $T_m$ value of biocomposite sample A is 136.49°C, while the $T_m$ value of the biocomposite sample B is 136.26°C. For sample B, the $T_m$ is only slightly lower than that of sample A. The insignificant differences of $T_m$ values of samples A and B clearly implied that there is no specific interaction between ChCl/Gly with matrix, explaining the decrease in mechanical and morphological properties of biocomposite sample B. Whereas, the $T_m$ value of biocomposite sample C is 134.23°C, exhibiting that the $T_m$s of the samples A and B are higher from that of sample C. This result suggests that the introduction of ChCl/Gly-HTAB into biocomposite affect the $T_m$ of sample by decreasing its value. On the other hand, the $T_m$ values of samples D and E are 133.42 and 133.18°C, respectively. It can be seen that the increase in loading of ChCl/Gly-HTAB has decreased values of the $T_m$ of the biocomposites. The decrease considerably could be related to the presence of interactions between both polymer and filler in the biocomposites, and it is also the indication of an improvement in composite compatibility.[45] Moreover, this has enhanced the soft segment of biocomposites and imparted greater toughness. This is reflected in the mechanical testing results where the impact and extension properties are increased compared to those without ChCl/Gly-HTAB. It is clearly seen in DSC thermograms that the compatibility of the HDPE/agar biocomposites introduced with the ChCl/Gly-HTAB is significantly improved. From the results obtained, the interactions have played an important role in decreasing the thermal properties of biocomposites. This was also supported by the TGA results, given in the next section.

## TGA characterization

The TGA thermograms of the HDPE/agar biocomposites with various contents of ChCl/Gly-HTAB are illustrated in the Figure 7. The initial decomposition (at 10% weight loss) and maximum decomposition temperatures values of the biocomposites are listed in the Table 3. The biocomposite sample A decomposes in three steps. In contrast, the biocomposites samples B to E decompose in four steps, showing the existence of additional decomposition steps in TGA thermograms after introduction of ChCl/Gly or ChCl/Gly-HTAB in the biocomposites. First decomposition step for biocomposite sample A corresponds to thermal decomposition of agar,[18] whereas second and third decomposition steps are believed to correspond to thermal decomposition of HDPE.[46] For biocomposites samples B to E their first and second

**Table 2.** Melting point temperature ($T_m$) of the HDPE/agar biocomposites as obtained from DSC thermograms.

| Sample | $T_m$ (°C) |
|--------|-----------|
| A | 136.49 |
| B | 136.26 |
| C | 134.23 |
| D | 133.42 |
| E | 133.18 |

DSC: differential scanning calorimetry; HDPE: high-density polyethylene.

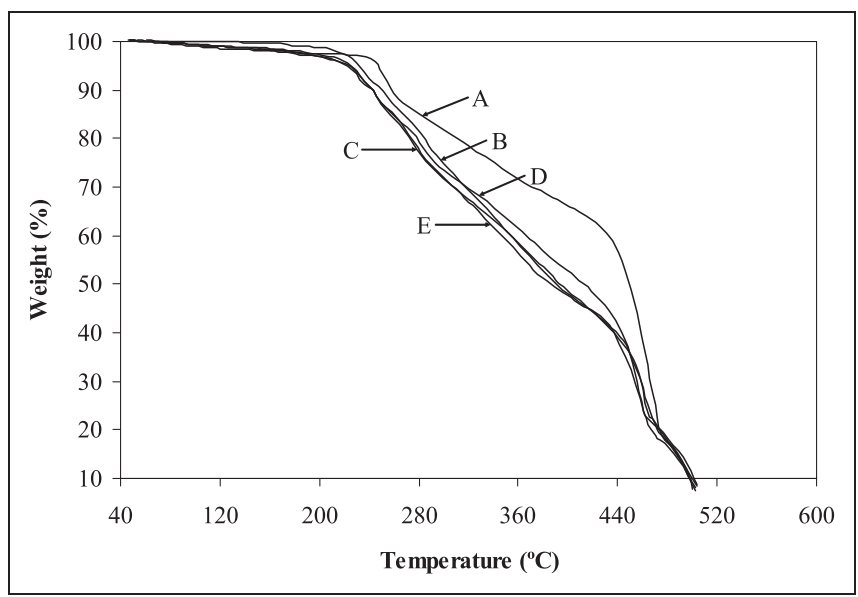

**Figure 7.** TGA thermograms of the HDPE/agar biocomposites.

decomposition steps are due to thermal decomposition of ChCl/Gly or ChCl/Gly-HTAB and agar, respectively, while third and fourth decomposition steps typically belong to thermal decomposition of HDPE. The initial decomposition temperature value of biocomposite sample A is 257.83°C whilst, the initial decomposition temperature values of biocomposites samples B, C, D and E are 250.52, 243.28, 243.18 and 243.09°C, respectively. It is clearly seen that initial decomposition temperature of the biocomposite sample B decreases with introduction of ChCl/Gly. A similar trend has been observed for the biocomposites samples C, D and E introduced with the ChCl/Gly-HTAB. This is as expected since the initial decomposition temperatures of ChCl/Gly and ChCl/Gly-HTAB are slightly lower compared to the other components in the biocomposites.[37] Furthermore, increase of the soft segment could

also make the initial decomposition temperature of the biocomposites decrease.[47]

On the other hand, the maximum decomposition temperatures values of biocomposites samples A, B, C, D and E are 468.29, 464.14, 456.06, 455.96 and 453.92°C, respectively. It is seen that the maximum decomposition temperature decreases almost linearly with the initial decomposition temperature. However, the maximum decomposition temperatures of samples C, D and E are much lower than that of the corresponding samples A and B. This might be caused by the presence of interactions between agar fillers and HDPE matrix, which effectively decreases the thermal decomposition of the overall biocomposites.[18] The maximum decomposition temperature of the HDPE/agar biocomposites is also decreased with an increase in the ChCl/Gly-HTAB content. It is obviously displayed that the thermal decomposition of the biocomposites depends on the loading of the component that can compatibilize the other components. This is due to the fact that HTAB has the lowest thermal decomposition temperature in comparison with the other components in the biocomposites. Thus, in terms of thermal stability, biocomposites samples C to E were found to be less stable than biocomposites samples A and B. This produces the biocomposites with low thermal dissociation characteristic that easily decomposed thermally.

### FTIR characterization

The FTIR spectra of the HDPE/agar biocomposites with several ChCl/Gly-HTAB contents are demonstrated in Figure 8, whereas FTIR bands of the biocomposites have been exhibited in the Table 4. The broad bands in the

**Table 3.** The initial and maximum decomposition temperatures of the HDPE/agar biocomposites acquired from TGA thermograms.

| Sample | Initial decomposition temperature (°C) | Maximum decomposition temperature (°C) |
|--------|----------------------------------------|----------------------------------------|
| A | 257.83 | 468.29 |
| B | 250.52 | 464.14 |
| C | 243.28 | 456.06 |
| D | 243.18 | 455.96 |
| E | 243.09 | 453.92 |

HDPE: high-density polyethylene; TGA: thermal gravimetric analysis.

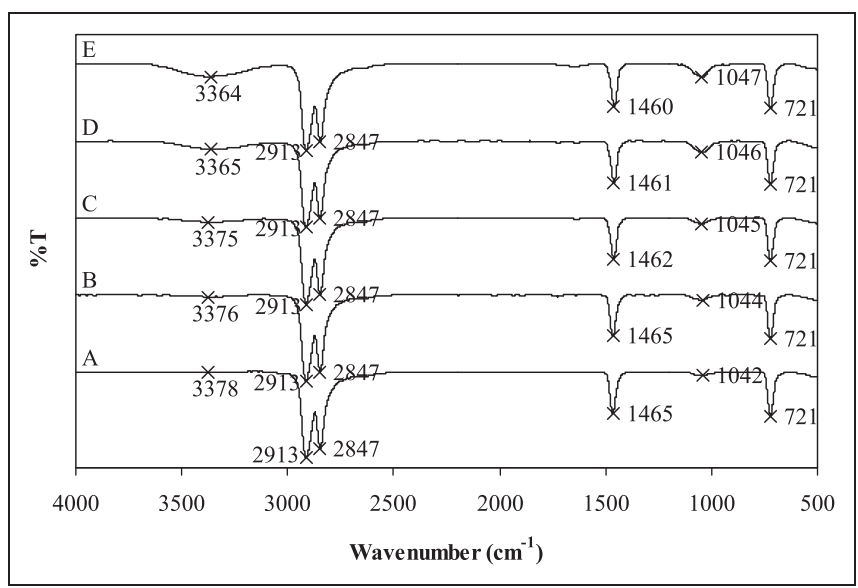

**Figure 8.** FTIR spectra of the HDPE/agar biocomposites with several ChCl/Gly-HTAB contents.

region of 3378 to 3364 cm$^{-1}$ from very weak to medium intensities could be assigned to the O–H stretching vibration of the alcohol group.[18,36] The noticeable bands with strong intensity at 2913 cm$^{-1}$ and 2847 cm$^{-1}$ are, respectively, attributable to the CH$_2$ asymmetric stretching and CH$_2$ symmetric stretching of the methylene group. The strong intensity of the bands at around 1465 to 1460 cm$^{-1}$ resulted from the C–H bending deformation.[48] The bands from very weak to medium intensities that are associated with C–O stretching of the alcohol group were found at approximately 1047 to 1042 cm$^{-1}$,[11,36] while the bands with medium intensity at 721 cm$^{-1}$ that were also present in all the prepared biocomposites revealed the CH$_2$ rocking deformation.[48]

The FTIR bands indicated that the wavenumbers of the biocomposites vary between each other. It is because the ChCl/Gly-HTAB has been introduced into the biocomposites. The bands of the O–H stretching of the alcohol group of the biocomposites samples C to E have significantly shifted to the lower wavenumbers regions compared to the one without HTAB such as sample B. This is owing to the formation of hydrogen bonding between the polar hydroxyl groups on the surfaces of agar fillers with the polar hydroxyl groups of the ChCl/Gly-HTAB. A parallel behavior has been perceived for the bands of C–H bending of methylene group of the biocomposites samples C to E where they have also considerably shifted to the lower wavenumbers regions compared with the biocomposite sample B. This is because the non-polar alkyl groups of the ChCl/Gly-HTAB have the capacity to interact with the non-polar groups of HDPE matrix through the hydrophobic-hydrophobic interaction.[30,49] Therefore, these results indicate the presence of physical interactions between agar-ChCl/Gly-HTAB and ChCl/Gly-HTAB-HDPE, which could provide the interaction link between the biocomposites components like agar-ChCl/Gly-HTAB-HDPE. This contributes to the increased compatibility of the biocomposites components. The results explain the increase in stabilization torque as well as mechanical and morphological properties of the biocomposites. On top of that, C–O stretching of the alcohol group shifted towards higher wavenumbers as the content of ChCl/Gly-HTAB increased in the biocomposites. This phenomenon was caused by the decrease in internal hydrogen bonding in the agar,[36] where strength of hydrogen bonding decreased from samples C to E. Although polar cation groups of HTAB molecules can form interaction with polar hydroxyl groups of agar fillers via surfactant binding (ion-dipole force),[32] the polar hydroxyl groups of the ChCl/Gly-HTAB have more affinity to form hydrogen bonding with the polar hydroxyl groups of agar fillers. In addition, the intensity of the bands of both O–H and C–O stretching of the alcohol group is also considerably increased with concomitant increasing of ChCl/Gly-HTAB contents. This is also due to the increase of interactions between HDPE matrix and agar fillers. In spite of this, there is no appearance of additional bands observed in the FTIR spectra when the ChCl/Gly-HTAB was introduced into the HDPE/agar biocomposites. Hence, the components of the biocomposites do not have any significant change in their chemical structures after melt processing with ChCl/Gly-HTAB.

## XRD characterization

The XRD patterns of the HDPE/agar biocomposites with different amounts of ChCl/Gly-HTAB are shown in the Figure 9. In general, all XRD patterns for each composition have a similar pattern where they possess two main diffraction peaks at 2θ values of 21.6° and 24.0°, which both were reflections of 110 and 200 planes of PE, respectively.[50] However, from the results it was found that the biocomposite sample A exhibited the highest intensity characteristic with one sharp peak at 2θ = 21.6°, indicating the higher crystallinity of the biocomposite than the other biocomposites. High crystalline properties in sample A was usually considered as ordered structure of HDPE chain segments in the biocomposite. While, in the biocomposite sample B with

**Table 4.** FTIR bands of the HDPE/agar biocomposites.

| Sample | Wavenumber (cm$^{-1}$) | | | | | |
| --- | --- | --- | --- | --- | --- | --- |
| | O–H stretching | CH$_2$ stretching | CH$_2$ stretching | C–H bending | C–O stretching | CH$_2$ rocking |
| A | 3378 | 2913 | 2847 | 1465 | 1042 | 721 |
| B | 3376 | 2913 | 2847 | 1465 | 1044 | 721 |
| C | 3375 | 2913 | 2847 | 1462 | 1045 | 721 |
| D | 3365 | 2913 | 2847 | 1461 | 1046 | 721 |
| E | 3364 | 2913 | 2847 | 1460 | 1047 | 721 |

HDPE: high-density polyethylene; FTIR: Fourier transform infrared.

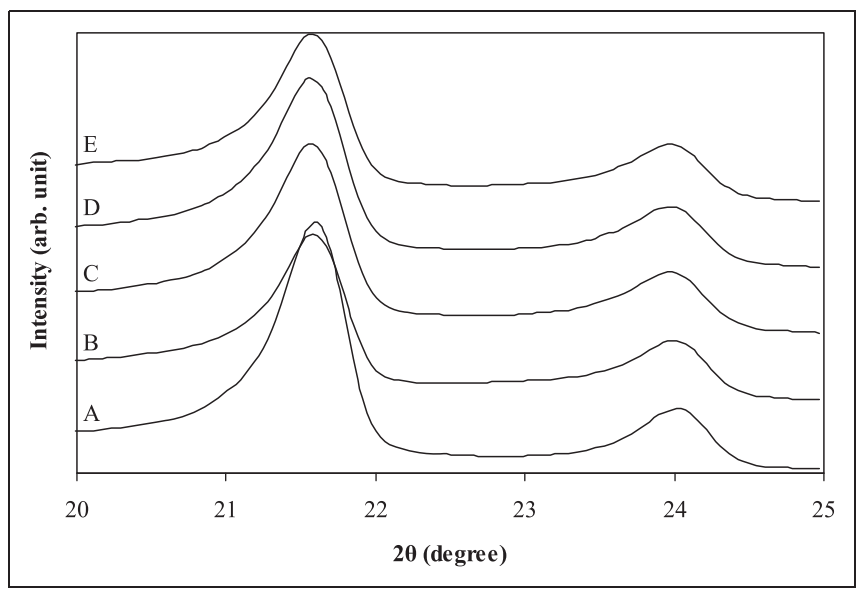

**Figure 9.** XRD patterns of the HDPE/agar biocomposites with different amounts of ChCl/Gly-HTAB.

introduction of the ChCl/Gly, the peak intensity at $2\theta = 21.6°$ is slightly decreased. Nevertheless, after introduction of the ChCl/Gly-HTAB the peak intensity of biocomposite sample C is significantly reduced. In fact, for the biocomposites samples D and E, their peaks intensities at $2\theta = 21.6°$ is lower than the peaks of the biocomposites samples A and B. Such diminution happened when the amount of crystalline phases in biocomposites decreased. Therefore, the introduction of ChCl/Gly-HTAB not only decreases the melting point and thermal decomposition temperatures but also declines degree of crystallinity of the biocomposites.

Moreover, the decrease in intensity of the peaks of XRD patterns at $21.6°$ is also due to the occurrence of interactions between HDPE and agar in biocomposites. If there were no interactions between those components, the diffraction peaks should remain sharp and well-defined.[51] From this result, the significant change in their peaks confirms that interactions have occurred between the components of biocomposites. This also exhibited that the crystallization of HDPE chain segments in biocomposites were almost constrained as a consequence of interactions between the biocomposites components. Thus, the crystalline phase of the biocomposites samples C to E are decreased compared with the biocomposites samples A and B. On the other hand, the peaks at $2\theta = 24.0°$ that can be observed in all the biocomposites samples nearly remained unchangeable and could be disregarded. In addition, none of the XRD patterns of the biocomposites samples showed any new isotropic peaks. In other words, the above results show that the ChCl/Gly-HTAB merely forms physical interactions (hydrogen bonding and hydrophobic-hydrophobic interaction) instead of strong chemical bonds (covalent bonding and ionic bonding) in compatibilizing of HDPE/agar biocomposites.

## Conclusions

From this study, the introduction of ChCl/Gly-HTAB has increased the stabilization torque of the HDPE/agar biocomposites. This certainly could be related to the increase of the compatibility between the components of biocomposites and results in increase of their melt viscosity. Moreover, the impact strength and tensile extension of the biocomposites introduced with ChCl/Gly-HTAB are also increased as exhibited in the mechanical testing results. The highest impact strength and tensile extension were found at the sample C with loading of 0.2 wt% HTAB. This reflects better interactions of biocomposites with ChCl/Gly-HTAB than biocomposites with ChCl/Gly and without ChCl/Gly-HTAB at a similar HDPE/agar loading. The number of agar fillers pull-out, melting point and thermal decomposition temperatures of the biocomposites are significantly decreased with the introduction of the ChCl/Gly-HTAB as indicated in the SEM, DSC and TGA results. These clearly exhibit improvement of the matrix-filler interfacial adhesions between HDPE and agar. Furthermore, the presence of physical interactions such as hydrogen bonding and hydrophobic-hydrophobic interaction in the biocomposites is also due to the introduction of the ChCl/Gly-HTAB as implied in the FTIR spectra and XRD patterns. This contributes to the increase of compatibility between both HDPE and agar. It can therefore be concluded that the eutectic-based ionic liquid containing surfactant, ChCl/Gly-HTAB, could compatibilize HDPE/agar biocomposites.

## Funding

This work was supported by the Ministry of Education Malaysia under Exploratory Research Grant Scheme (ERGS) (project code: ERGS/1-2013/5527185).

## Acknowledgements

The authors thank the staffs of Laboratory of Biocomposite Technology, Institute of Tropical Forestry and Forest Products, Universiti Putra Malaysia, and the School of Chemical Sciences and Food Technology, Faculty of Science and Technology, Universiti Kebangsaan Malaysia, for the technical assistance and the facilities.